\documentclass[letter]{spie}
\pdfoutput=1
\usepackage{graphicx}
\usepackage{graphics}
\usepackage{color}
\usepackage{SIunits}
\DeclareGraphicsExtensions{.pdf,.png,.jpg}

\title{Digital frequency domain multiplexing readout electronics for
  the next generation of millimeter telescopes}

\author{Amy N. Bender\supit{a}, Jean-Fran\c{c}ois Cliche\supit{a}, Tijmen de Haan\supit{a}, Matt A. Dobbs\supit{a,b}, Adam J. Gilbert\supit{a},  Joshua Montgomery\supit{a}, Neil Rowlands\supit{c}, Graeme M. Smecher\supit{d}, Ken Smith\supit{c}, Andrew Wilson\supit{c}
\skiplinehalf
\supit{a}Department of Physics, McGill University, 3600 Rue University, Montreal, Quebec H3A 2T8, Canada 
\skiplinehalf
\supit{b}Canadian Institute for Advanced Research,
CIFAR Program in Cosmology and Gravity, Toronto, ON, M5G 1Z8, Canada
\skiplinehalf
\supit{c}COM DEV Canada, 303 Terry Fox Drive, Ottawa, ON, K2K 3J1, Canada
\skiplinehalf
\supit{d}Three-Speed Logic, Inc., Vancouver, BC V6A 2J8, Canada
}

\authorinfo{Send correspondence to A. N. Bender: amy.bender@mail.mcgill.ca}

\begin{document}

\def\mk{$mK$}
\def\microk{$\muK$}
\def\tcmb{$T_{CMB}$}
\def\vbias{$V_{b}$}
\def\mk{$mK$}
\def\microk{$\muK$}
\def\tcmb{$T_{CMB}$}
\def\vbias{$V_{b}$}

\maketitle

\begin{abstract}
Frequency domain multiplexing (fMux) is an established technique for the readout of transition-edge sensor (TES) bolometers in millimeter-wavelength astrophysical instrumentation. In fMux, the signals from multiple detectors are read out on a single pair of wires reducing the total cryogenic thermal loading as well as the cold component complexity and cost of a system. The current digital fMux system, in use by POLARBEAR, EBEX, and the South Pole Telescope, is limited to a multiplexing factor of 16 by the dynamic range of the Superconducting Quantum Interference Device pre-amplifier and the total system bandwidth.  Increased multiplexing is key for the next generation of  large format  TES cameras, such as SPT-3G  and POLARBEAR2, which plan to have on the of order 15,000 detectors. 

Here, we present the next generation fMux readout, focusing on the warm electronics.  In this system,  the multiplexing factor increases to 64 channels per module (2 wires) while maintaining low noise levels and detector stability.  This is achieved by increasing the system bandwidth, reducing the dynamic range requirements though active feedback, and digital synthesis of voltage biases with a novel polyphase filter algorithm.   In addition, a version of the new fMux readout includes features such as low power consumption and radiation-hard components making it viable for future space-based millimeter telescopes such as the LiteBIRD satellite.
\end{abstract}

\section{Introduction}

Technological advances in millimeter-wavelength detectors over the past decade have resulted in telescopes with sensitivity limited mainly by the number of the detectors.   The next generation of large millimeter-wavelength telescopes, such as SPT-3G \cite{benson2014} and POLARBEAR2\cite{tomaru2012}, will have on the order of 15,000 detectors, an order of magnitude larger than current cameras.   Millimeter cameras employ bolometers that  operate at sub-Kelvin temperatures to detect the incident radiation.  Reading out each bolometer individually would create a significant thermal load on the cryogenic stages of the camera.  Additionally, the complexity and cost of such a system would be restrictive.   In order to accommodate the large number of bolometers, the readout is multiplexed, a technique in which many detectors are read out on a single pair of wires.  

The bolometer cameras currently operating on the South Pole Telescope \cite{austermann2012} and POLARBEAR telescope \cite{kermish2012} as well as the EBEX balloon-boorne telescope\cite{grainger2008} use digital frequency domain multiplexing (fMux) \cite{dobbs2012} to read out arrays of transition-edge sensor (TES) bolometers.  In those systems, up to 16 bolometers can be multiplexed together by separating the operating bandwidth of each detector in frequency space.  In this paper, we present the next generation of the digital fMux  readout in which 64 bolometers are multiplexed together, focusing on the room temperature electronics.  In addition to the 64x digital fMux electronics for ground-based telescopes, we present a second version of the electronics designed for satellite platforms.

\section{System Overview}
\label{sec:overview}
A TES bolometer absorbs heat from incident photons, which results in a change in the electrical resistance of a thermistor. Applying a strong voltage bias to the thermistor provides electrical power, holding the TES in the superconducting transition.  The constant voltage bias enables negative electrical feedback, where the electrical power changes with the TES resistance keeping the total power constant, and changes the current through the device\cite{irwin1995}.    

\begin{figure}[htb]\centering
\includegraphics[width=1\textwidth]{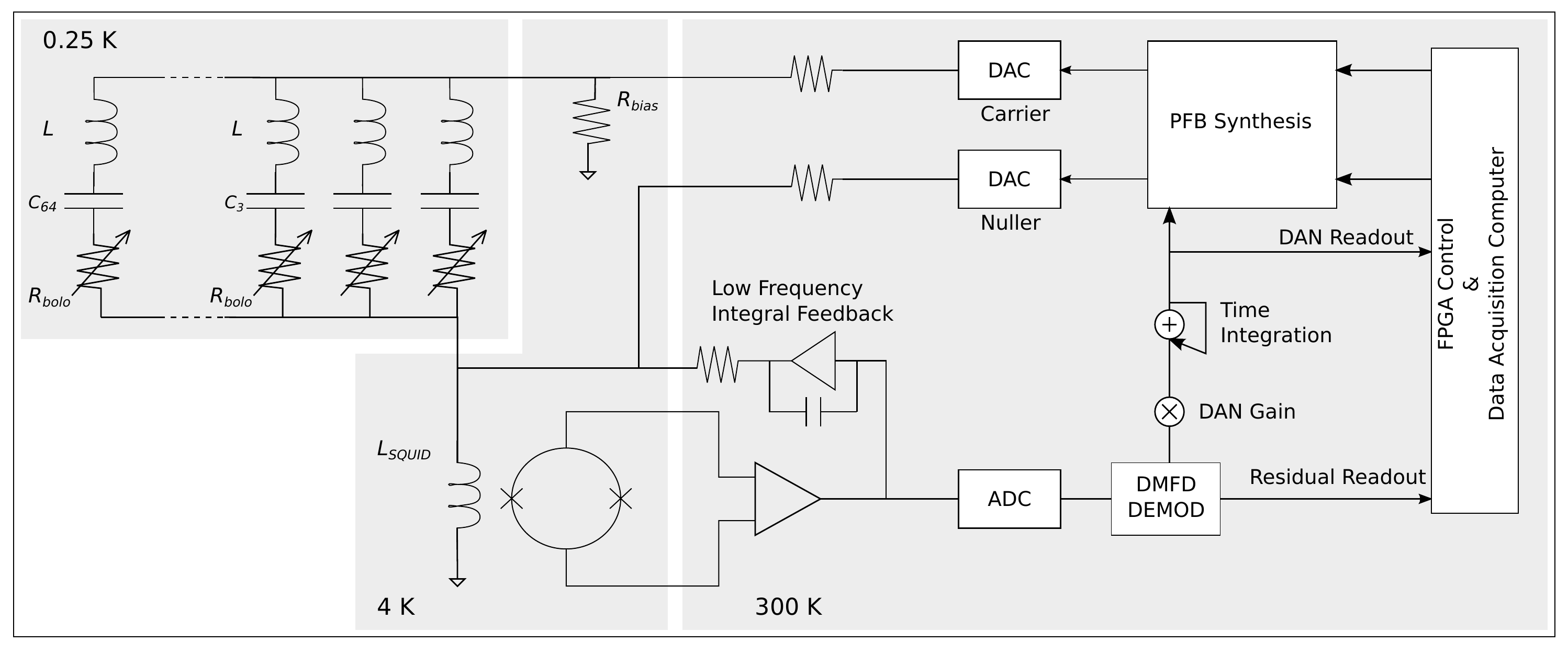}
\caption{Schematic overview of the 64 channel digital frequency domain multiplexing readout system.}
\label{fig:fmuxschematic}
\end{figure}

In an fMux readout, bolometers are multiplexed together by placing each in series with an inductor $L$ and capacitor $C$ creating a unique frequency resonance (see Figure \ref{fig:fmuxschematic}).  The $LCR_{\mathrm{bolo}}$ segments are placed in parallel with each other. A sinusoidal voltage bias is tuned to the resonant frequency of each bolometer and summed together into a single waveform (referred to as the \textit{comb of carriers}) and input on a pair of wires. The sky signal amplitude modulates each carrier and appears as sidebands in the resulting current. Because the sidebands for each bolometer occupy a unique frequency band, the signals from each of the bolometers are transmitted over a single pair of wires. 

The summed bolometer signals are amplified using Superconducting Quantum Interference Devices (SQUIDs). The SQUID is a low-noise, low input-impedance transimpedance amplifier.  After the SQUIDs, the signal undergoes additional stages of amplification and signal processing in the room temperature electronics before being digitized and demodulated down to baseband.

One significant challenge in increasing the number of bolometers multiplexed together is dynamic range.   As the multiplexing factor increases the dynamic range of the SQUID quickly becomes a limiting factor.
Since the sky signal is modulated into the carrier sidebands, the SQUID dynamic range requirements can be significantly reduced by injecting an inverted copy of the carrier comb with a 180 degree phase shift prior to the SQUID input coil (the \textit{nuller}).  In the legacy version of the digital fMux readout, statically applying the nuller was sufficient to ensure that the remaining current into the SQUID coil met the dynamic range requirement.  However, the static nulling method is no longer sufficient when the multiplexing factor increases to 64. Instead, an active feedback loop referred to as Digital Active Nulling (DAN)\cite{dehaan2012} nulls both the carriers and sidebands for each bolometer.

The dynamic range of the synthesizer chain also limits the number of channels that can be multiplexed together.
A typical bolometer for a ground-based telescope has a saturation power of 15~pW, and will require a voltage bias of about 3~$\micro \volt_{\mathrm{RMS}}$ for $R_{\mathrm{bolo}} = 1 \ohm$, corresponding to a current of 3~$\micro \ampere_{\mathrm{RMS}}$.  
The achievable dynamic range at the bolometer from the synthesizer chain is derived from the properties of the digital-to-analog converter (DAC), a current source, paired with the requirement that the total noise contribution of the DAC at the bolometer be no more than the SQUID noise (3 $\rm{pA}/\sqrt{\rm{Hz}}$).  For the new electronics, the upper limit of the dynamic range of the synthesizer chain at the bolometers is 1.2~$\milli \ampere_{\mathrm{P-P}}$ (424~$\micro \ampere_{\mathrm{RMS}}$). 

In the simplest conceptual mode of operation all synthesized carrier sinusoids are started in phase, limiting the multiplexing factor to far below the goal of 64. By randomizing the relative phases of each carrier frequency, the available bandwidth can be used far more effectively. 

The crest factor (sometimes called the ``peak-to-average ratio'') is the ratio of the maximum excursion of a waveform amplitude to the root-mean-square (RMS) waveform amplitude,
\begin{equation}
  \mbox{Crest\ Factor} = \frac{A_{\mathrm{Peak}}}{A_{\mathrm{RMS}}} .
\end{equation}
The crest factor for a single sinusoid is $A_{\mathrm{Peak}}/A_{\mathrm{RMS}}=\sqrt{2}$.  Crest factor minimization refers to the practice of reducing the crest factor through a judicious choice of waveform parameters, in this case, the carrier phases. For $N$ carriers of peak amplitude $A$, the total waveform RMS grows proportional to $\sqrt{N}$. In the worst-case scenario, such that all carriers are in phase, the waveform peak amplitude would be $N \cdot A$, and the crest factor would be $N \cdot A/  \left(\frac{\sqrt{N}\cdot A}{\sqrt{2}}\right)=\sqrt{2N}$. An algorithm to adjust the phase of many sinusoids and search for a set of parameters that mininimizes the crest factor exists\cite{Lindeman2012}. Since the DAN feedback technique is constantly making small adjustments to the phases of each nuller we look for smoothness of the parameter-space to ensure the existence of broad minima instead of a narrow global minimization. We note that no simple analytic formula to determine a minimum crest factor exists. Instead, the problem is usually addressed with numerical techniques wherein a set of phases is chosen randomly, and the crest factor is estimated by simulating the waveform.
We find that, when the phases are chosen randomly, the typical crest factor is 4.25 for 64 carriers, with a spread of $\sim 4\%$. The resulting waveform peak is $4.25\sqrt{N}\cdot A/\sqrt{2}$, approximately 3$\sqrt{N}$ larger than the maximum value for a single sinusoid.

Given these results, we can relate the required dynamic range $A_{\mathrm{dr,P-P}}$ to the peak amplitude of a single sinusoid, 
 \begin{equation}
\label{eqn:MaxCalcICE}
\frac{A_{\mathrm{dr,P-P}}}{2} =  3\sqrt{N}\cdot A, 
\label{eqn:dynamicrange}
\end{equation}
 where $A = \sqrt{2} \cdot 3 \micro \ampere_{\mathrm{RMS}}$ for the desired voltage bias.
For ground-based bolometers with a multiplexing factor of 64 a minimum DAC dynamic range of $A_{\mathrm{dr,P-P}}  =204~\micro \ampere_{\mathrm{P-P}}$ is required.  The typical dynamic range of the new 64x digital fMux electronics is 1.2 $\milli \ampere_{\mathrm{P-P}}$, a factor of 6 higher than the minimum required.


Overall, there are three key components of the 64x multiplexing digital fMux readout: SQUID operation, the DAN feedback, and the generation of sinusoidal voltage biases.   The following sections describe these components, focusing on the changes necessary to increase the multiplexing factor from 16 to 64.

\begin{figure}[htb]\centering
\includegraphics[width=0.45\textwidth]{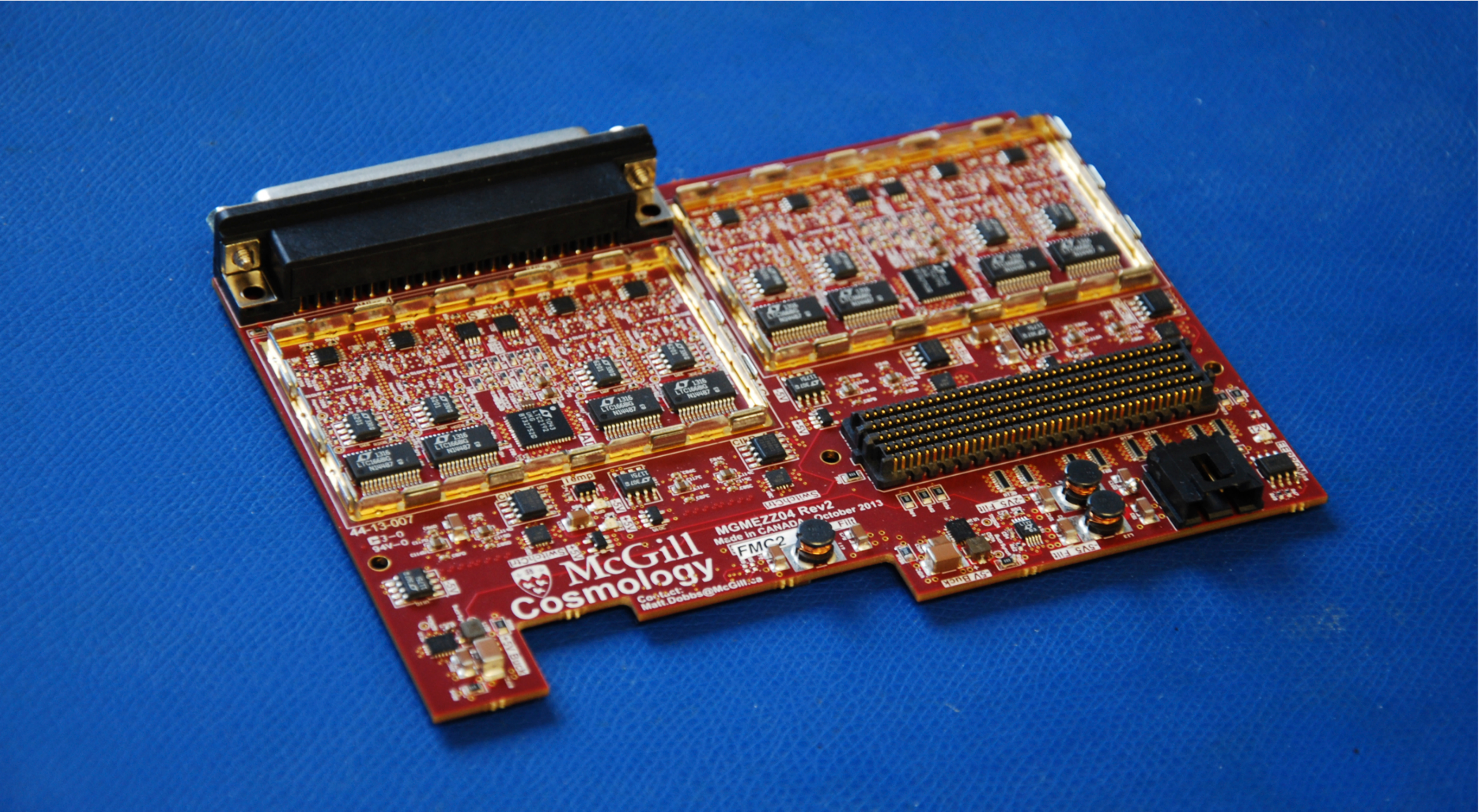}
\includegraphics[width=0.45\textwidth]{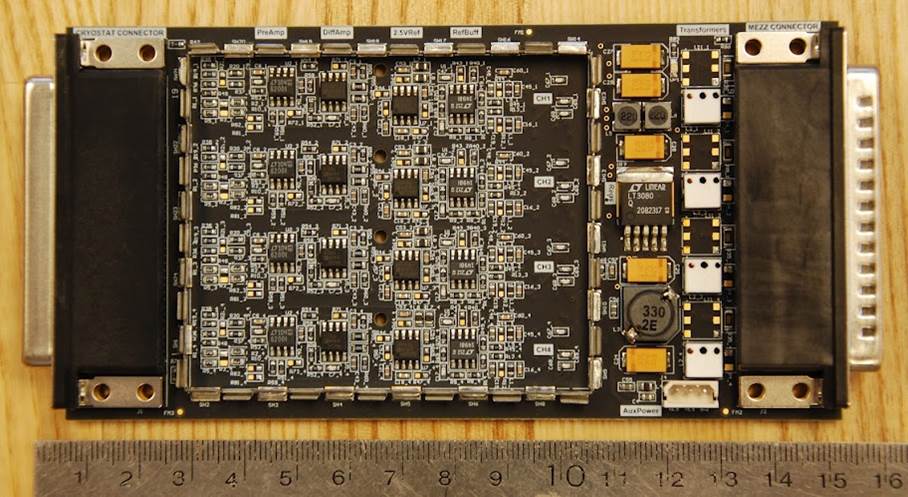}
\caption{Photographs of the synthesizer/digitizer mezzanine board (left) and SQUID controller board (right) for the 64x digital multiplexing readout system.}
\label{fig:cotsphotos}
\end{figure}

\subsection{SQUID Operation}

The high gain, low input-impedance, and low noise of SQUIDs make them ideal devices for the first amplification stage of the fMux readout. When a current bias of sufficient magnitude is applied to a SQUID it develops a periodic voltage output as a function of the magnetic flux in the input coil. By applying a direct current flux bias to the input coil, the SQUID can be tuned to the region of maximum gain and linearity (see Figure \ref{fig:vphi}). This functionality, together with other tuning controls, is supplied by the SQUID controller board (Figure \ref{fig:cotsphotos}), each of which can operate four SQUIDs.

The legacy digital fMux readout included a shunt feedback loop\cite{dobbs2012} to lock the SQUID at the desired tuning point and limit the input signal to the region of linear sensitivity. The 64x digital fMux employs DAN feedback and therefore no longer requires the SQUID shunt feedback. The shunt feedback had a bandwidth of 1~MHz, limiting the multiplexing factor. Removing the shunt feedback, we are able to substantially increase the bandwidth of the system to 10~MHz or higher.  

While DAN feedback ensures operation in the linear region of the SQUID, it is desirable to also have a low-frequency feedback loop to ensure the SQUID stays locked at the optimal tuning point.  For example, SQUIDs can be sensitive to changing external magnetic fields (e.g., when rotating through the Earth's field).  We implement an analog integral feedback loop between the output of the first stage amplifier and the SQUID input coil (see Figure \ref{fig:fmuxschematic}) that has a bandwidth of approximately 10~kHz to ensure a fixed point of operation on long time scales with no additive noise.

\subsection{DAN}

Digital Active Nulling (DAN) provides feedback to the input coil of the SQUID over a small frequency range around each of the carrier frequencies \cite{dehaan2012}.  In DAN, the amplitude and phase of the nuller current is continuously adjusted with digital signal processing algorithms implemented in firmware at rates much faster than the bolometer time constant. Samples are taken from the demodulator, weighted by a user programmable gain and accumulated to provide integral control to each of the 64 channels. The sky signal is now encoded in the nuller feedback amplitude and the SQUID output becomes an error signal that is used to adjust the nulling levels.  

DAN was first implemented as a modification to the legacy digital fMux readout. The South Pole Telescope polarization camera (SPTpol) successfully used DAN for the 2013 observing season, with no significant change in the measured white noise level of the bolometers.

\subsection{Digital Signal Processing Firmware}

The signal path of the digital fMux system has evolved significantly since the most recent published technical description \cite{smecher2012}. These changes have allowed the digital fMux system to aggressively scale to higher channel counts, without substantial increases in either power consumption or Field Programmable Gate Array (FPGA) resource usage.  Additionally, we have shifted development of our hardware, firmware, and software stack to a new FPGA motherboard based on Xilinx's Kintex-7 product line. This platform is a major upgrade from our legacy Virtex-4 platform, and grants us substantial opportunities to improve the firmware's performance and design margins. In the following sections, we describe these changes.

Although portions of the signal path (e.g., the synthesizer) have become wholly unrecognizable from their earlier incarnations, we have maintained consistent control and data interfaces. The model from the legacy digital fMux readout\cite{smecher2012} still serves as a conceptual guide, even though it no longer maps closely to the hardware implementation.
The digital signal processing (DSP) framework will continue to evolve. In particular, recent optimizations in the synthesizer structure will also be applied to the demodulator structure, reducing the complexity and power requirements associated with the readout portion of the system. In addition, we continue to optimize the signal path to fully exploit the capabilities of our new hardware platform.

\subsubsection{ICE Motherboard FPGA Platform}

McGill's legacy digital fMux readout platform was based on a custom Virtex-4 motherboard first commissioned in 2008 \cite{dobbs2008}. FPGA technology has progressed rapidly since then. To reduce cost, increase channel density, and take advantage of newer FPGAs, we have developed and built a new family of readout hardware. This family includes backplane/crate hardware, FPGA motherboards, and daughtercards based on the FPGA Mezzanine Card (FMC) standard.  In this section, we provide a brief technical description of the CPU and FPGA subsystems on the new ICE motherboard.

The CPU subsystem is a stand-alone ARM Cortex-A8 processor, connected to 1~GB DDR3 SDRAM and dual gigabit Ethernet ports. The CPU manages the many small sensors (temperature, voltage) and interfaces (mezzanine power switches, backplane control, various SPI, USB, and I2C buses). It communicates with the FPGA via high-speed SPI, JTAG, and a PCIe link. The CPU boots from a network or from an SD flash card, and manages the FPGA's boot-up and command interactions.

The FPGA subsystem centers around a Xilinx Kintex-7 '420T FPGA. The FPGA's 28 high-speed serial (GTX) ports are committed to two QSFP ports, one selectable SFP or PCIe port, and a 19-channel backplane interconnect. The bulk of the FPGA's I/O is committed to hosting two high-pin-count (HPC) FPGA Mezzanine Card (FMC) sockets for daughtercards.

Like the previous FPGA platform, the ICE motherboard uses only a single clock reference for all of its on-board systems. This clock is typically sourced from a backplane, although it can also be generated on-board or supplied via an SMA connector. Single-sourced clocking is critical to avoid aliased noise from contaminating the analog data.

\begin{figure}[htb]\centering
\includegraphics[width=1\textwidth]{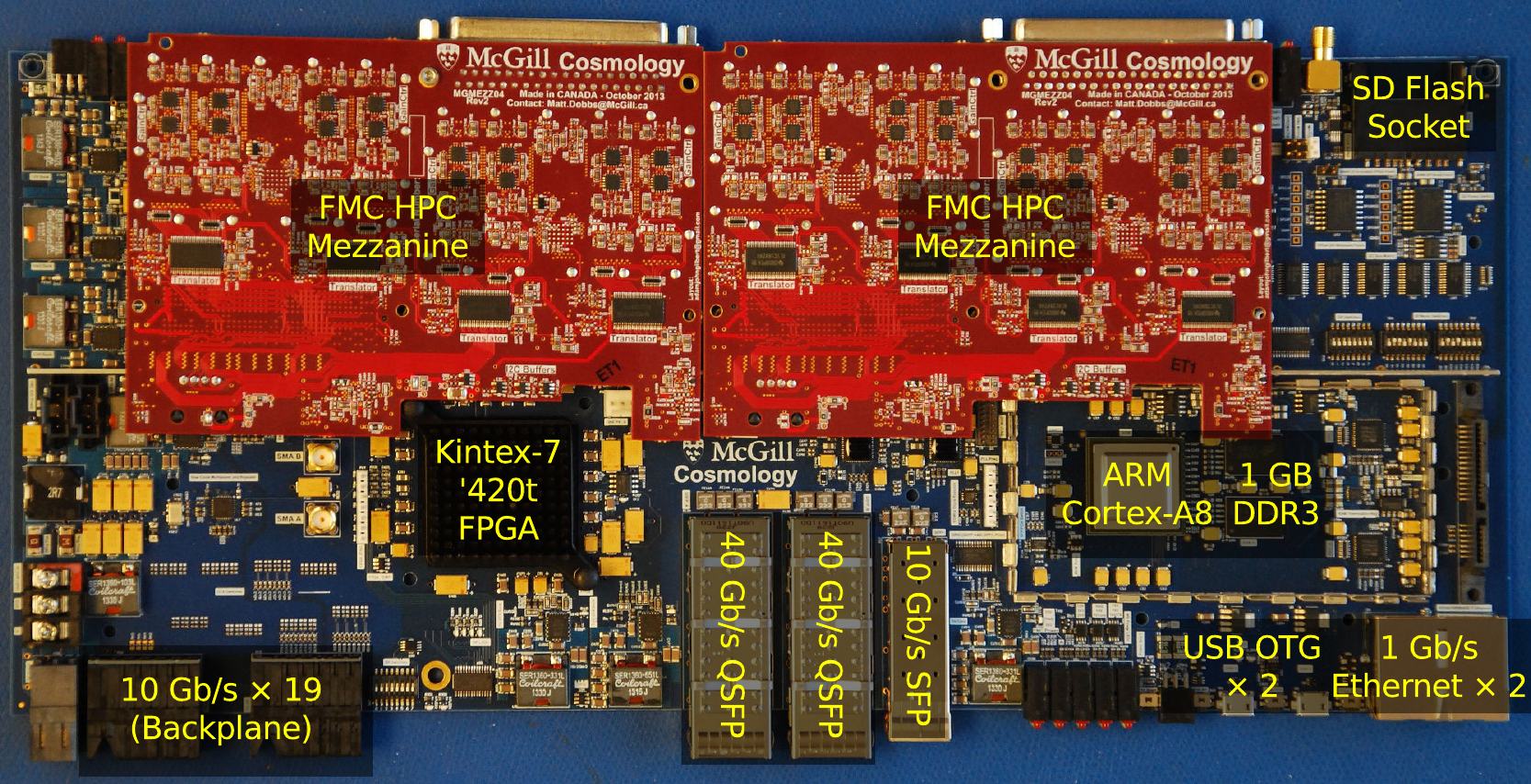}
\caption{The ICE motherboard, McGill's new FPGA platform. The black heat sink covers the FPGA; the CPU, DDR3 RAM, and Ethernet PHYs are visible on the right. The red mezzanines are two FMC-compliant high-speed data-acquisition boards.}
\label{fig:iceboard}
\end{figure}

\subsubsection{Mezzanine Interface}
The mezzanine board  (see Figure \ref{fig:cotsphotos}) interfaces between the ICE FPGA motherboard and SQUID controller board.  
The mezzanines host the following subsystems:
\begin{itemize}
\item Digital-to-Analog Converters (DACs) for synthesizing carrier and nuller waveforms,
\item Analog-to-Digital Converters (ADCs) for digitizing bolometer signals,
\item A low-speed digital interface for communicating with SQUID controller boards, and
\item Programmable-gain amplifiers (PGAs) for optimizing ADC and DAC dynamic range.
\end{itemize}
These mezzanines, intended for deployment in ground-based experiments, each support 4 SQUID readout modules.

\subsubsection{Synthesizer}

\begin{figure}[htb]\centering
\includegraphics[width=1\textwidth]{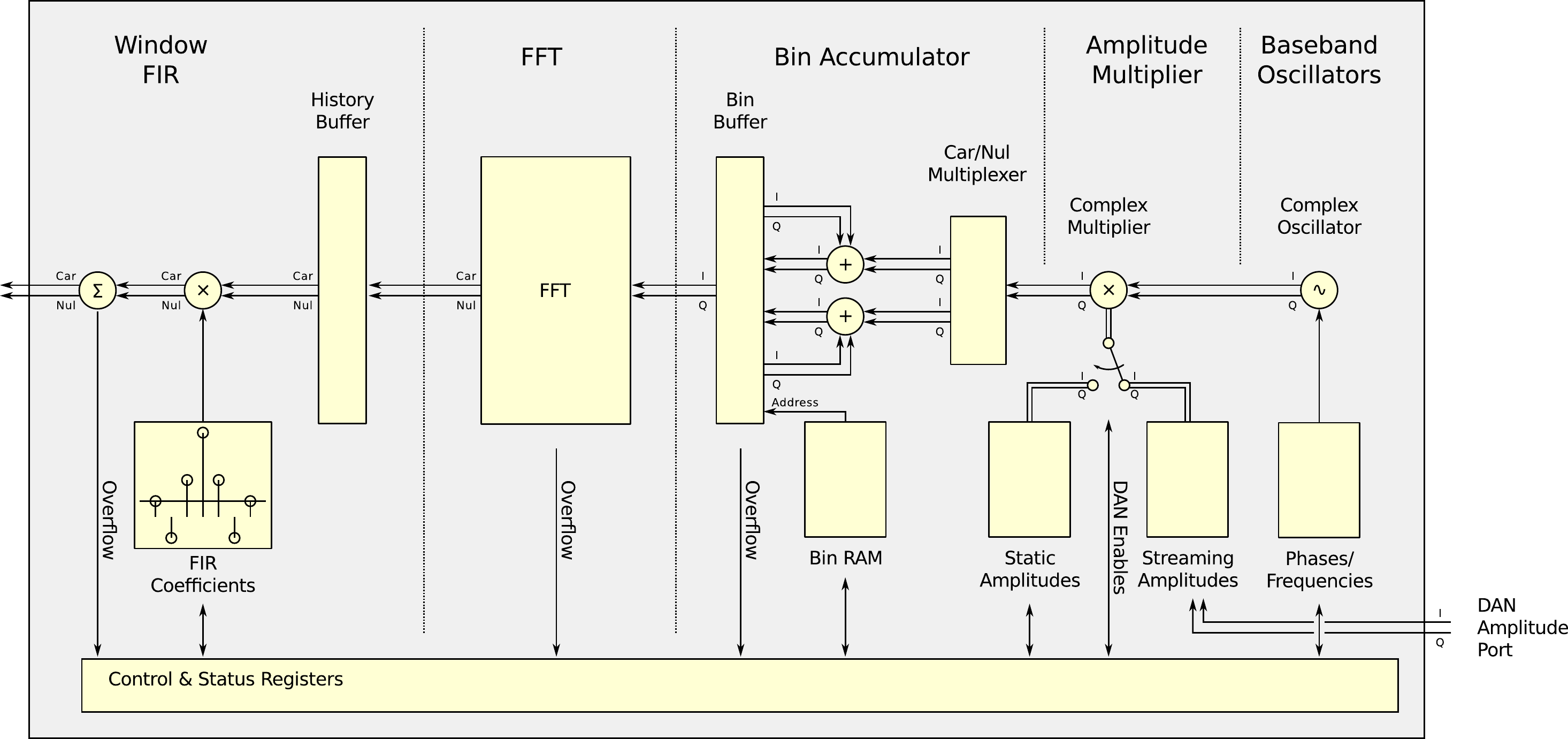}
\caption{Polyphase Filter Bank (PFB)-based synthesizer.}
\label{fig:pfb}
\end{figure}

Of all the portions of the DSP logic, the synthesizer module (Figure \ref{fig:pfb}) has evolved the most. It is barely recognizable from its earlier form\cite{smecher2012}. In the following section, we describe the current synthesizer design.
Formerly, the carrier and nuller synthesizers directly modulated signals from baseband to the relevant carrier/nuller frequency using a Direct Digital Synthesis (DDS) oscillator and mixer. Both the oscillator and mixer operate at the system's output sampling rate (25~MSPS). Since the FPGA logic is capable of running substantially faster than 25~MHz, the synthesizer implementation has been optimized by time-multiplexing each mixer and DDS among eight channels. The system's scalability is ultimately linear; doubling the number of channels required approximately double the logic.

The new carrier and nuller synthesizer is based on a Polyphase Filter  Bank (PFB) synthesizer \cite{crochiere1983}. This synthesizer is best described backwards: The DAC output bandwidth (operating at 20~MHz) is divided into 128 bins; each synthesizer channel frequency is reduced to a choice of the nearest bin, plus some residual frequency offset. The offset is applied via a DDS mixer, which operates at a low enough sampling rate that a single time-multiplexed DDS and mixer suffices for more than 128 channels.

In logical order, the synthesizer's signal path begins with the offset mixer. After generating sinusoids for each channel at a low sampling rate (625~kHz), amplitudes are applied from either static settings (i.e. in non-DAN mode) or a streaming amplitude port (i.e. in DAN mode.)

Next, a multiplexer combines carrier and nuller in such a way that, after the Fast Fourier Transform (FFT), the carrier appears solely in the real (“I”) outputs, and the nuller appears solely in the imaginary (“Q”) components.  This permits the use of a single, complex FFT instead of two real FFTs. This multiplexing process is also responsible for combining channels that share the same frequency bin, translating the system from a time-multiplexed set of individual channels to a single FFT input buffer.
Next, the FFT processes blocks of input data into blocks of output data.

Finally, a polyphase Finite Impulse Response (FIR) filter applies a windowing function that smoothly combines adjacent blocks of FFT outputs. The result is a 20~MSPS stream of independent carrier and nuller data, which is sent to the appropriate DACs. The carrier comb provides the detector biases, and the nuller approximately cancels the signal at the SQUID.

\subsubsection{Demodulator}

\begin{figure}[htb]\centering
\includegraphics[width=1\textwidth]{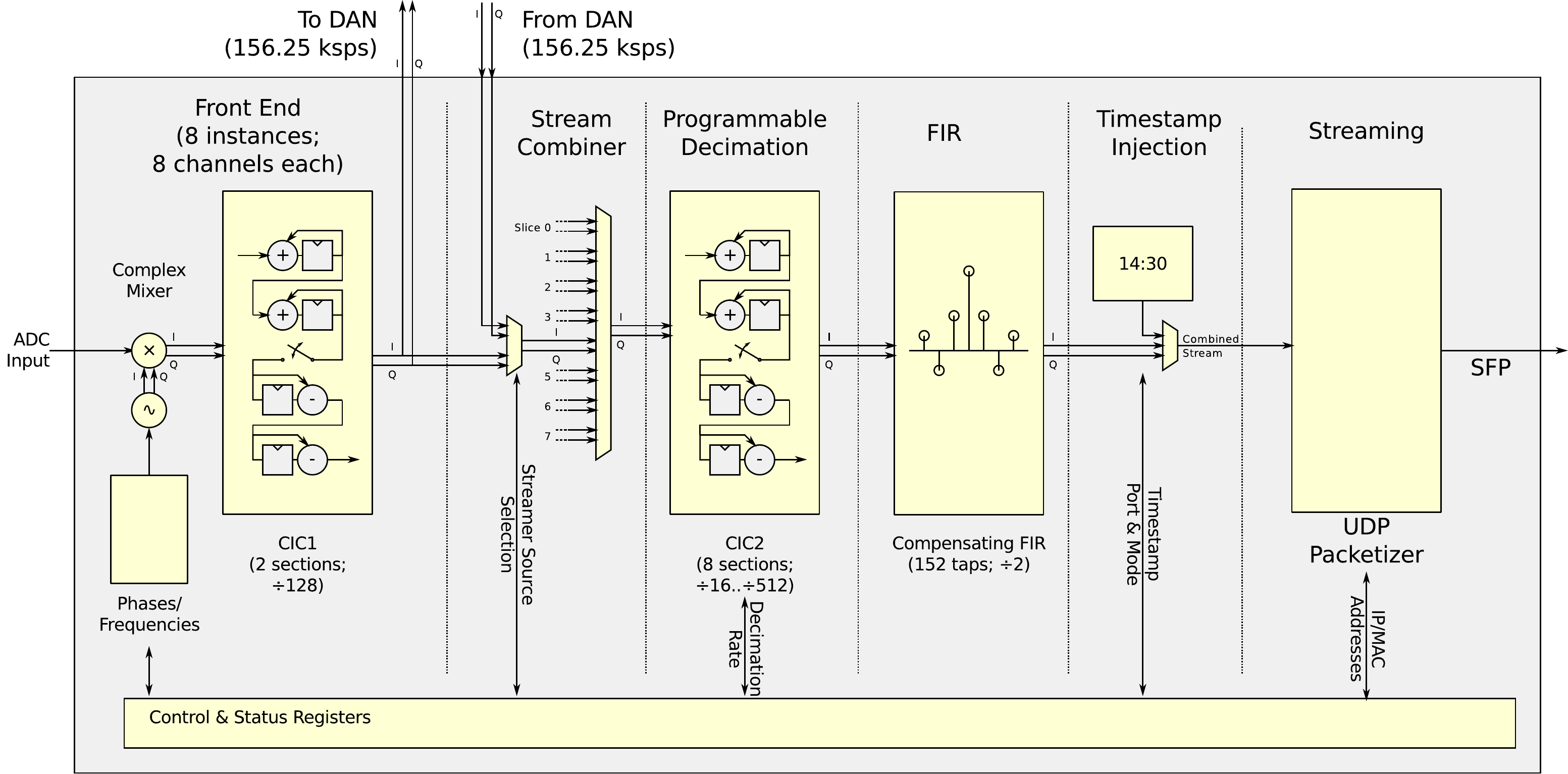}
\caption{The demodulator signal path, shown for a single SQUID readout module. The digitized SQUID signal is first mixed to baseband, then decimated by a CIC filter. In DAN mode, the signal is then passed to a dedicated DAN module for processing. Next, it is decimated by a second CIC filter, processed by a compensating FIR filter, timestamped, and streamed across the network.}
\label{fig:demod}
\end{figure}

The demodulator design (Figure \ref{fig:demod}) has evolved only slightly from the previous implementation\cite{smecher2012}. It still consists of a chain of DDS oscillators, mixers, Cascaded Integrator-Comb (CIC) filters, and FIR filters.

The digitized inputs from the system's ADCs are first processed by DDS-based complex mixers, creating 64 channels at different frequencies from a single ADC source.

Once mixed down to baseband, the signal in each channel is decimated by a factor of 128 using a CIC filter (CIC1). The output of this first decimation step is passed from the demodulator module out to the external DAN module, which automatically adjusts nuller amplitudes when the DAN feedback loop is closed.

The decimated output from the CIC1 filter is also passed to an 8-stage, 112-bit CIC filter (CIC2). This filter is designed with a variable decimation rate, selectable between ÷16, ÷32, ÷64, ÷128, ÷256, and ÷512. The filter's extreme word length is required to gracefully handle these decimation rates without a degradation in signal quality. To keep CIC2's implementation tractable, the 112-bit words are broken down into 14-bit sub-words and computed across eight consecutive clock cycles.

Finally, the output from CIC2 is compensated using an 152-tap FIR filter. This filter is necessary to correct passband non-uniformity due to the CIC1 and CIC2 filters.

In the legacy DSP designs \cite{smecher2012}, CIC2 was a fixed-rate decimator, followed by 5 FIR filters. Streamed data could be chosen from any of these FIRs.  In contrast with the current design, the older version was extremely costly in terms of DSPs and block RAMs within the FPGA. Functionally, the two designs are indistinguishable.

\subsubsection{Streaming \& Control Interface}

The digital fMux control and streaming interfaces are evolving gradually from the legacy system\cite{smecher2012B}. In this section, we briefly describe recent improvements.

Most importantly, the system's software stack now runs on the ICE motherboard's Cortex-A8 CPU. This CPU is an order of magnitude faster than the older system's MicroBlaze CPU, and includes a hardware floating-point unit. The increase in processor speed takes a substantial burden off the system's software development.

Now that the CPU is physically separated from the FPGA, we cannot simply access the signal path as a memory-mapped peripheral. The control interface has been moved to an ARM controlled SPI link, which permits reads and writes of the FPGA registers.

For data streaming, the FPGA generates UDP packets directly, and emits them via its SFP interface to a gigabit Ethernet port.

\subsection{Cryogenic Components}
\label{sec:cryocomponents}

The previous sections summarized the room temperature electronics for the 64x digital fMux readout.  Parallel improvements to the cryogenic components (the SQUIDs, cables, inductors, and capacitors depicted in the 4~K and 0.25~K portions of Figure \ref{fig:fmuxschematic}) of the readout system are also underway.  These developments are presented in complementary proceedings \cite{barron2014, hattori2014}.  Here, we describe the cryogenic components used during the laboratory commissioning of the room temperature electronics presented in Section \ref{sec:performance}.

For the results presented in this paper, cryogenic readout electronics similar to those currently deployed with the SPTpol camera \cite{henning2012} are used.  A two-part manganin wire harness connects the SQUID controller board to a set of NIST series-array SQUIDs \cite{huber2001} inside the cryostat at a temperature of 4~K.  A stripline connects the SQUIDS to an SPTpol style LC board and a prototype SPTpol bolometer wafer. The bolometers are dark, i.e. isolated so that very little incident radiation is absorbed. The LC board is populated with chips of 22~$\micro$H inductors (also fabricated at NIST) and commercial capacitors.  The capacitors are selected to give $LCR_{\rm{bolo}}$ resonances equally spaced from 200~kHz -- 5~MHz for the 38 working bolometers on the wafer.  An additional six resonances are created by populating 1~$\Omega$ calibration resistors and the appropriate capacitors on the LC board.   

These cryogenic components have two important parasitic contributions that can affect the stability and noise of TES bolometers. The first is the equivalent series resistance (ESR) of the capacitors in the $LCR_{\rm{bolo}}$ filter. At frequencies less than 1~MHz, the ESR in the commercial capacitors is on the order of  0.1-0.3~$\Omega$, small in comparison to the bolometer normal resistance ($\sim 1.5~\Omega$). However, ESR increases as a function of frequency. When the ESR is comparable to the bolometer normal resistance, the TES is no longer able to operate as deeply in the superconducting transition. The second parasitic of interest is the stray inductance in series with the bias resistor and that of the stripline\cite{dobbs2012}.  These strays alter both the amplitude of the voltage bias and associated noise level at the bolometer.

\section{Flight Representative Electronics}

\begin{figure}[htb]\centering
\includegraphics[width=0.4\textwidth]{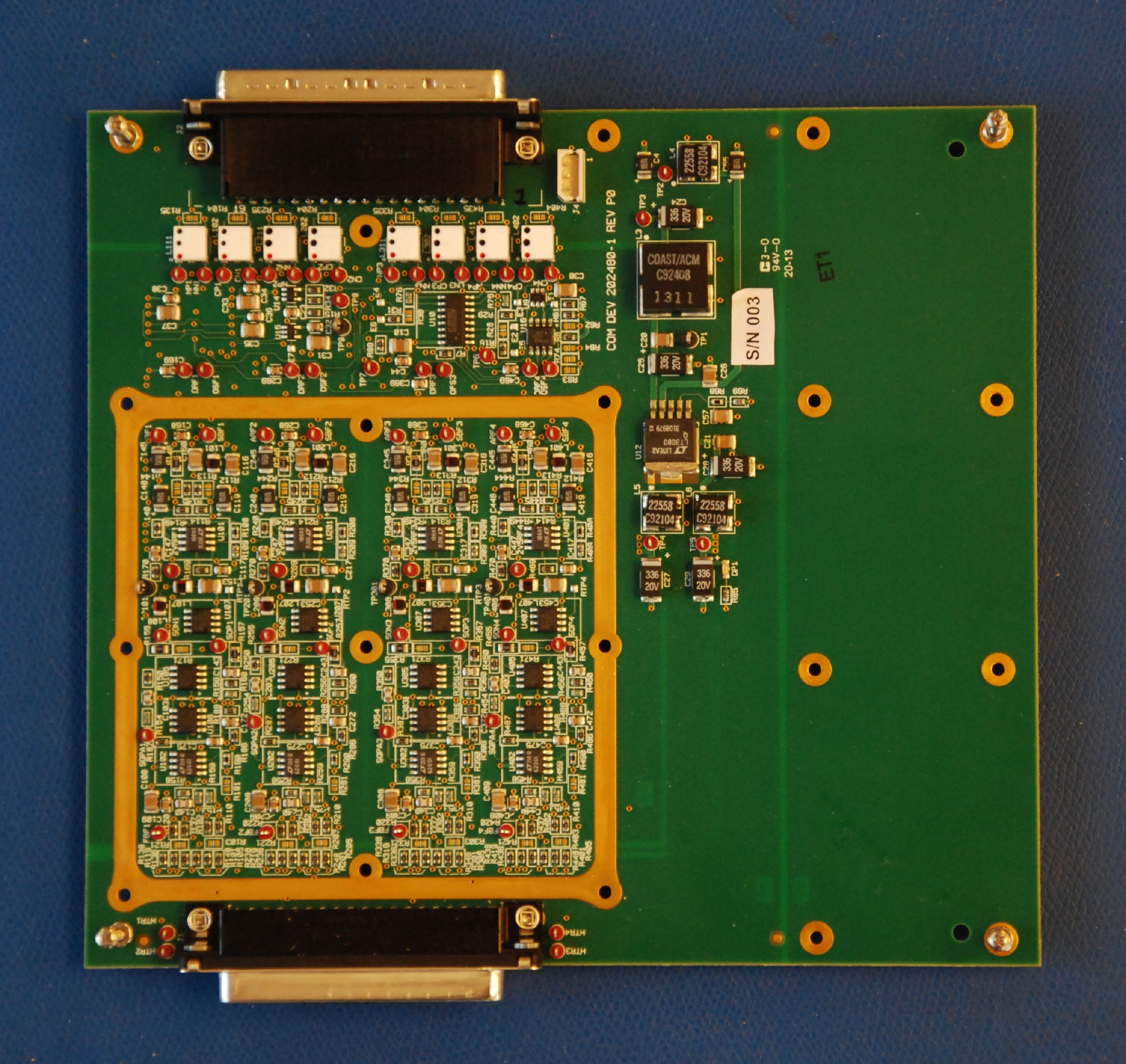}
\includegraphics[width=0.52\textwidth]{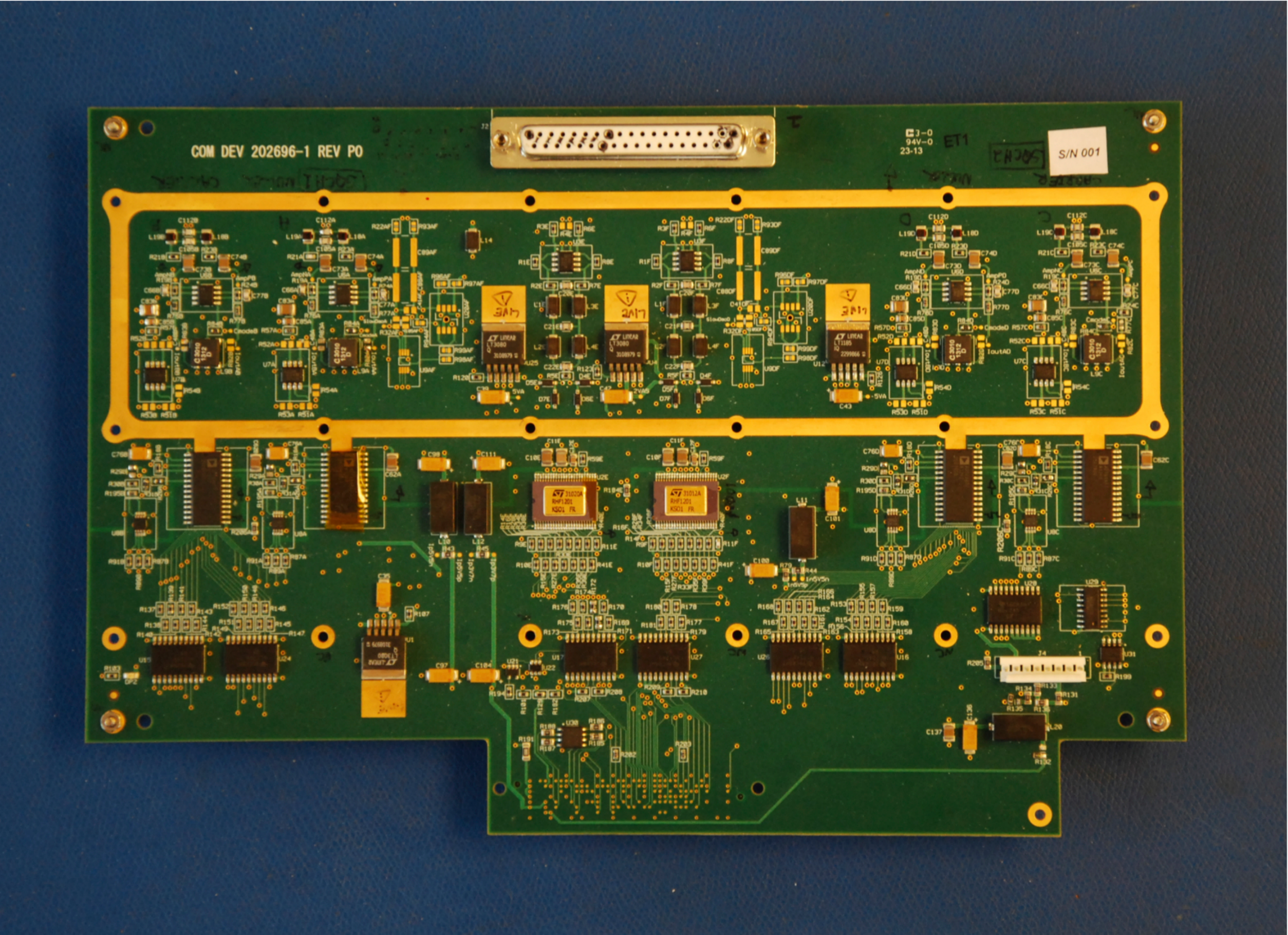}
\caption{Photographs of the the flight representative synthesizer/digitizer mezzanine board (left) and SQUID controller board (right).  The flight representative mezzanine and SQUID controller each support two and four modules, respectively.}
\label{fig:frphotos}
\end{figure}

TES detectors with SQUID-based multiplexed readout are an enabling technology for a new generation of space missions. These include, for example, the LiteBIRD mission \cite{hazumi2012} and the SAFARI instrument on the SPICA mission\cite{khosropanah2012}.  
A second set of room temperature electronics (mezzanine and SQUID controller boards) were developed in collaboration with COM DEV Canada to be flight representative and demonstrate the technological suitability for satellite platforms (see Figure \ref{fig:frphotos}). A flight representative FPGA motherboard has not yet been developed as FPGA technology evolves rapidly and the requirements of an FPGA motherboard for a satellite application would be highly mission-specific. Consequently, the commissioning of the flight representative electronics used a Xilinx Kintex-7 FPGA evaluation board (model KC705) for the digital signal processing platform.

The general architecture for the flight representative electronics is the same as that described in Section \ref{sec:overview}, but space applications have additional design requirements, such as reduced power consumption and robustness to the environment.  The digital fMux power budget is dominated by the fixed overhead in running the boards and the number of independent mux modules.  A substantial reduction in power consumption per bolometer is achievable by increasing the multiplexing factor to 64. Furthermore, the contribution of the generally power-hungry flight electronics components was offset by modifying the SQUID controller and part of the mezzanine design  to use single rail supplies. Further power savings were made by allowing the software to dynamically scale down the reference current of the high speed DACs, which also enables fine tuning of the system gains. Overall, with the KC705 board and four modules of 64 bolometers each, we measure a power consumption of 48.9~mW per bolometer, which satisfies the LiteBIRD design goal of 50~mW per bolometer\cite{litebird}. Further reductions are possible with higher multiplexing factors.

The Flight Representative SQUID controller and mezzanine board have also been redesigned to operate in the environment of space.  The first environmental consideration is radiation. Assuming a minimum mission lifetime of 2.5 years at the sun-earth L2 point, the electronics will experience a total ionizing dose of 50~krads under one millimeter of aluminum shielding. Electronic components with a radiation rating of 50-100~krads were selected. This required the use of the radiation-hardened Analog Device's AD768 DAC to generate the carrier and nuller signal at $>$25~MSPS. 
The electronic components of both boards were chosen based on existing suitable flight parts with matching commercial versions with the same performance specification, circuit topology, and die layout. In most cases, the commercial versions of the components have been used on the boards as they differ from the flight counterparts only in terms of packaging and screening. 
The layout of the boards includes the area necessary to accommodate the larger packages of flight components, ensuring representative part spacing.

The range of temperatures over which the electronics will operate is another important environmental consideration.
Assuming an approximately fixed attitude with respect to the Earth/Moon and the Sun and the thermal fluctuations in similarly oriented satellites, the digital fMux electronics should expect to experience slowly varying temperatures ($\sim 1^{\circ}$C per 24 hours) between $0^{\circ}$C and $+40^{\circ}$C. Flight qualified components typically have a range of storage temperatures much larger than this, so no special component selection is required. To ensure that the readout system performance is not degraded as a function of ambient temperature, the flight representative mezzanine and SQUID controller boards were operated in a temperature controlled chamber at temperatures between -$20^{\circ}$C to $+40^{\circ}$C (in order to show margin against the likely future requirements). 
Both power consumption and noise decrease by about $10\%$ as the temperature is reduced from $+40^{\circ}$C to $-20^{\circ}$C. The bias currents and references show excellent temperature stability and vary less than 1\% over the full temperature range, which does not impact system operation. Similarly, the frequency response of the electronics changes by a few percent and does not affect system operation. Overall, the system performs well across the full temperature range, with a small benefit to operating at temperatures below $0^{\circ}$C.  

A final environmental concern is the effect of electromagnetic interference (EMI) from the spacecraft electronics on the digital fMux electronics (specified by MIL-STD-461).  The analog portion of the electronics on both the mezzanine and SQUID controller boards are sensitive to EMI from the digital portion, therefore it is critical to self-shield (both in the ground-based and flight representative versions).  The flight representative boards include bolt-down machined aluminum EMI enclosures with EMI gaskets.

\section{System Performance}
\label{sec:performance}

In this section we present the performance of the 64x digital fMux readout electronics with the full system, including the laboratory cryogenic components. The results shown equally represent the ground-based electronics controlled with the ICE motherboard and the flight representative versions that were evaluated using the KC705 evaluation board.

\subsection{SQUID Operation}

\begin{figure}[htb]\centering
\includegraphics[width=0.55\textwidth]{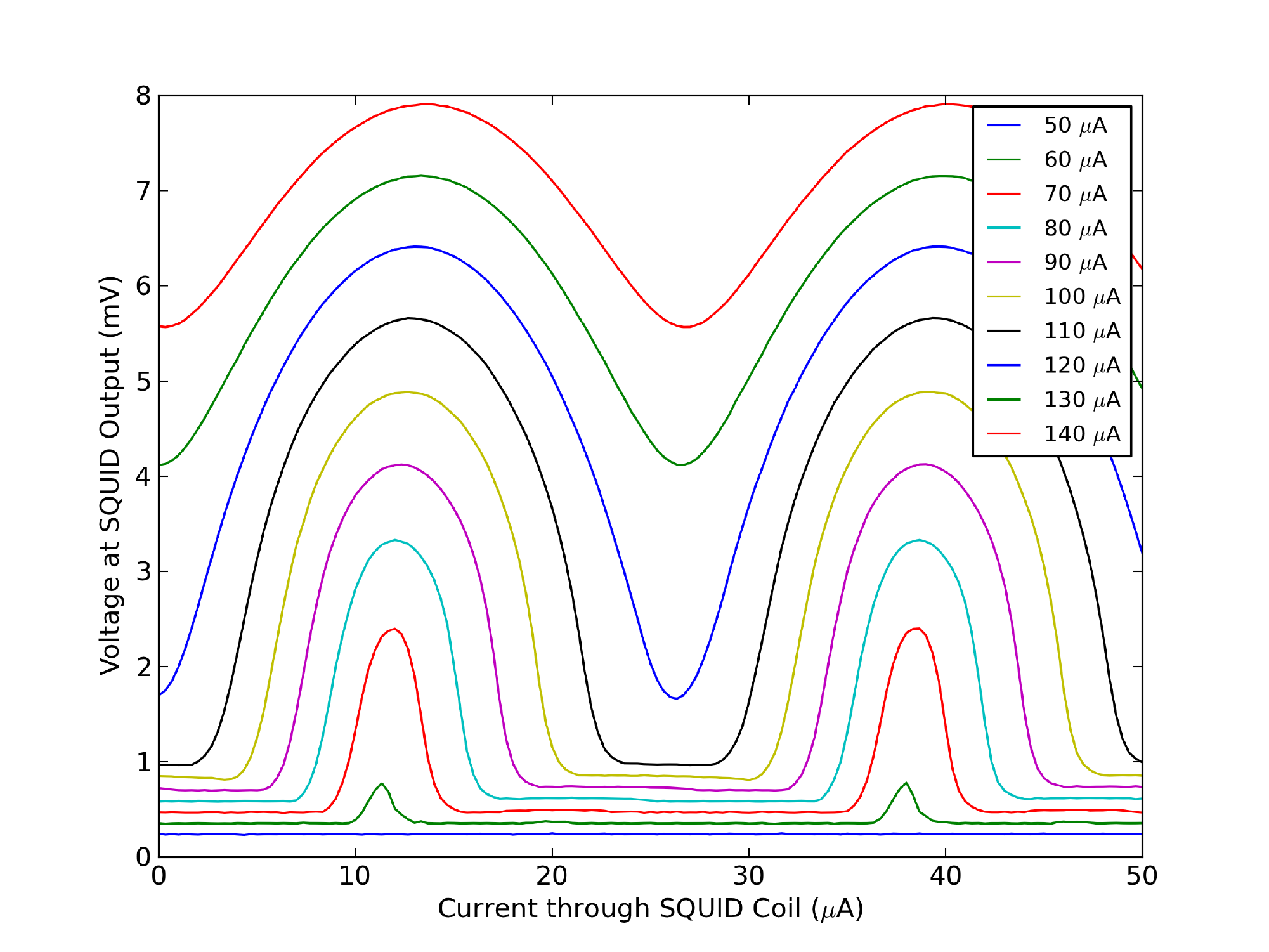}
\caption{SQUID voltage response as a function of current through the SQUID input coil for a range of bias current values.}
\label{fig:vphi}
\end{figure}

The performance of the SQUID controller is first evaluated on the benchtop. All DACs are exercised through a range of input values and the output voltage is measured.  To ensure a 10~MHz bandwidth, the frequency response of the board is also measured. After passing the benchtop tests, the SQUIDs in the cryostat are operated through a range of bias currents and flux bias currents (Figure \ref{fig:vphi}). The resulting voltage-flux curves have the expected sinusoidal shape and peak-to-peak amplitude.

Tuning the SQUID to the optimal bias point, we are able to measure the transimpedance (current to voltage gain) by injecting a small calibrated sinusoid on the nuller wires. Measuring the amplitude after demodulation, we find a transimpedance of approximately 383~$\Omega$. This value is typical for these laboratory quality SQUIDs, and is comparable to the transimpedance measured with the previous digital fMux readout.

We also characterize the behavior of the analog low-frequency integral feedback loop. To achieve this, the SQUID is first tuned to the point on the voltage-flux curve with the maximum peak-to-peak response. The integral feedback is enabled and a pilot signal is injected on the nuller wires. The flux bias DAC is then scanned through a range of values, mimicking the effect of a changing external magnetic field. We find that the amplitude of the pilot signal has a maximum deviation from its mean amplitude of less than $1\%$ while scanning through approximately 1/2 of the SQUID flux quanta. 

\subsection{Operation of 64 Channels}

\begin{figure}[htb]\centering
\includegraphics[width=0.9\textwidth]{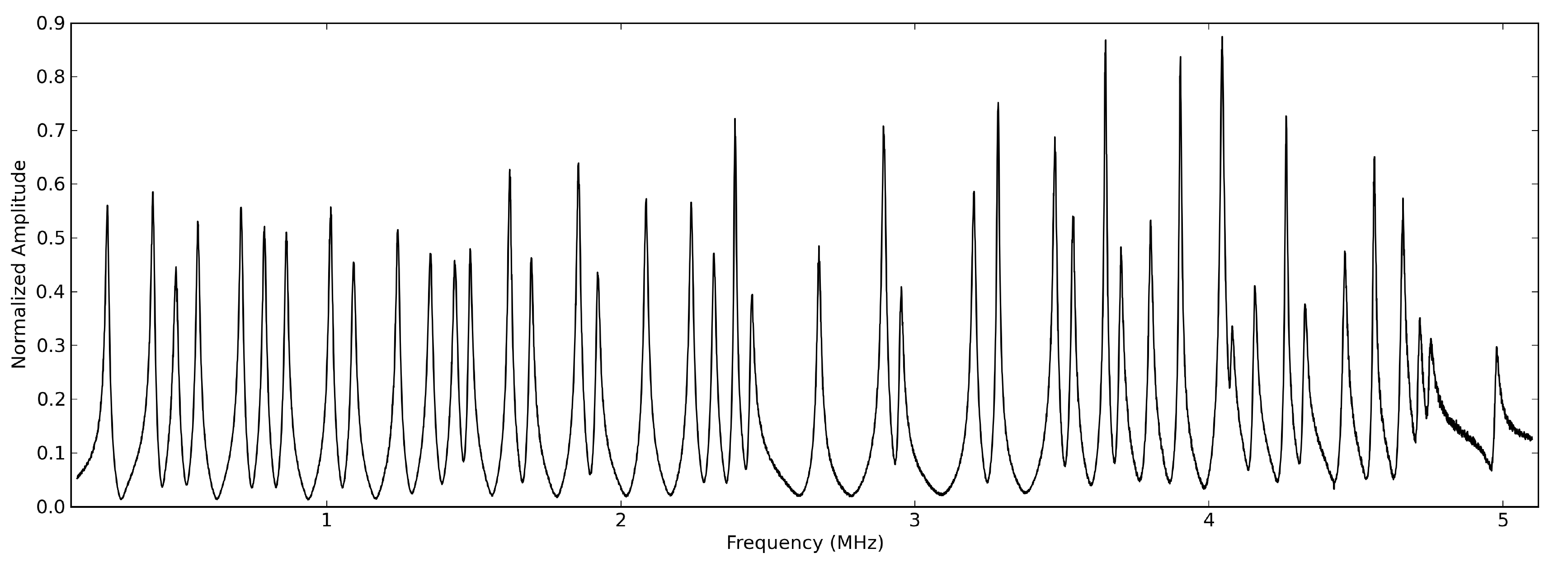}
\caption{Response as a function of frequency for a readout module containing 38 bolometers and six calibration resistors multiplexed together between 0 and 5~MHz.}
\label{fig:netanal}
\end{figure}

Before the bolometers can be operated in the superconducting transition, the specific location of the $LCR_{\rm{bolo}}$ resonances must be known. We measure the frequency response of the cold electronics by injecting pilot signals on the carrier and nuller wires and measuring the output demodulated amplitude (see Figure \ref{fig:netanal}). Signals are injected simultaneously using all 64 readout channels, each at a different frequency, iterating until the full bandwidth of the system is probed. As stated in Section \ref{sec:cryocomponents}, the module shown here has 38 bolometers and an additional six resistors for a total of 44 resonances. 

\begin{figure}[htb]\centering
\includegraphics[width=0.7\textwidth]{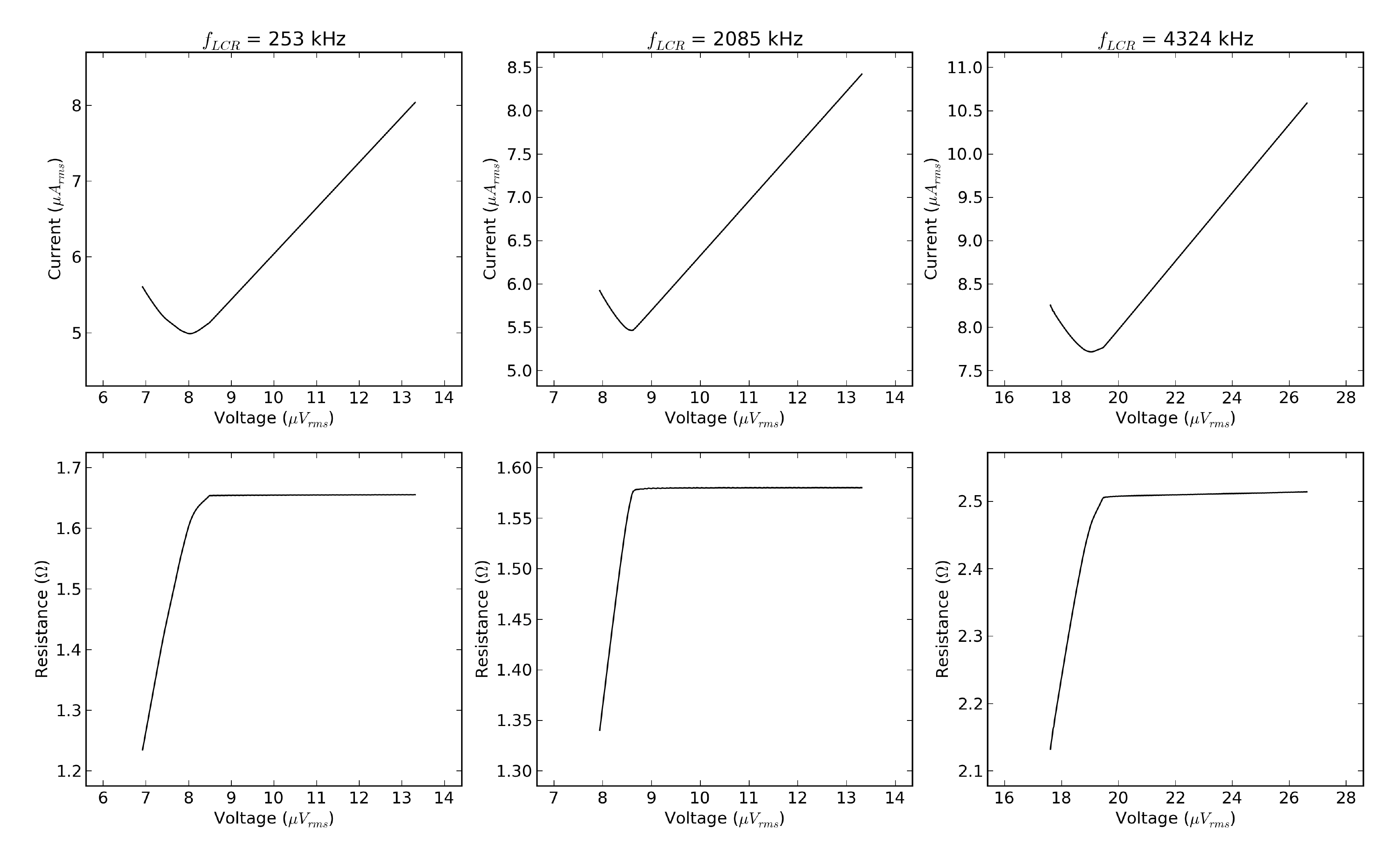}
\caption{Top row: Measured current as a function of voltage bias for three bolometers (low-, mid- and high-frequency resonance) as they are lowered into the superconducting transition. Bottom row: Bolometer resistance as a function of voltage bias for the same three bolometers.}
\label{fig:ivcurves}
\end{figure}

Bolometers are operated by first raising their physical temperature above that of the superconducting transition. At this temperature the bolometers behave like resistors. DAN feedback is enabled and voltage biases are applied to the bolometers (this is referred to as the \textit{overbiased} state). The voltage biases are large enough to hold the bolometers normal while the physical temperature is lowered back to the base value below the transition temperature. The voltage biases are simultaneously reduced for all bolometers, slowly lowering them into the superconducting transition. Figure \ref{fig:ivcurves} shows the measured current through the bolometer as a function of decreasing voltage bias for three bolometers at different carrier frequencies. The linear current-voltage region at high bias voltage shows the normal regime of the bolometer that then transitions through a turnaround and into the superconducting transition at low bias voltage.  Similarly, the resistance is shown to be approximately constant in the normal regime and then sharply decreases as the bolometer enters the superconducting transition.
  
We find a general trend that as carrier frequency increases the minimum stable fraction of the normal resistance at which the bolometer operates increases (it is biased higher in the superconducting transition). This is due to the known increase in capacitor ESR with frequency (see Section \ref{sec:cryocomponents}).  Despite this behavior, we are able to successfully operate the 38 bolometers in the superconducting transition. To more closely approximate the conditions for operating 64 bolometers, the remaining 20 channels are set to off-resonance frequencies.  These channels are used to simulate the total load on the nuller, carrier, and demodulator signal chains and can also be used to measure the properties of the readout electronics.   

\subsection{Noise Performance}

\begin{figure}[htb]\centering
\includegraphics[width=0.55\textwidth]{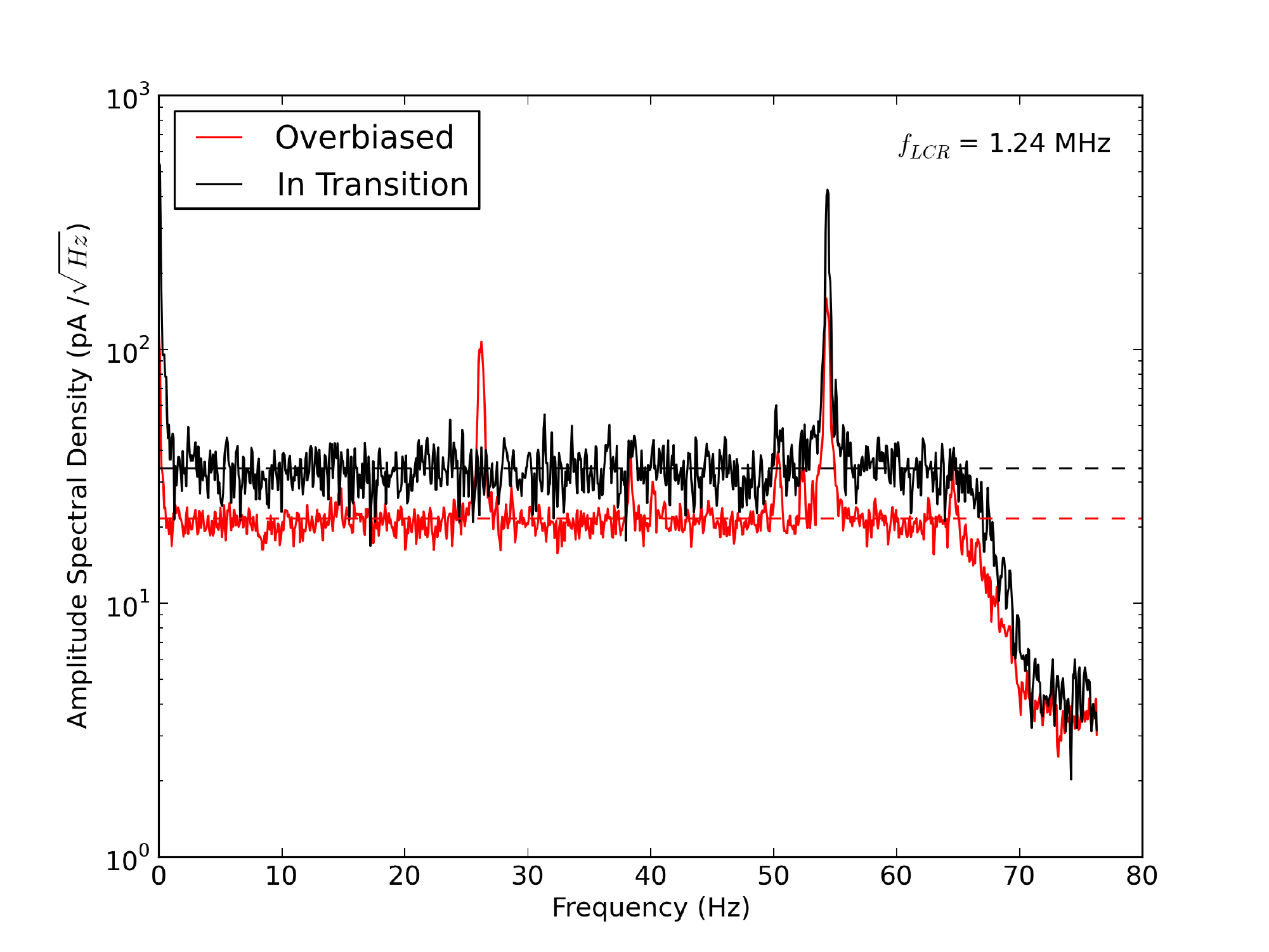}
\caption{Amplitude spectral density for a bolometer while overbiased (red) and while in the superconducting transition (black). A linear fit to the time domain data was removed prior to taking Fourier transform to remove a drift from the slowly changing physical temperature. Note that the white noise level (indicated by the dashed lines) increases as expected in the transition due to the phonon noise contribution.}
\label{fig:noisepsd}
\end{figure}

TES bolometers have been demonstrated to achieve photon-noise-dominated performance \cite{bock2009}. A key performance criterion for the 64x digital fMux readout is to contribute a negligible amount to total noise. We show the amplitude spectral distribution for a representative dark bolometer operated in the laboratory with a bias frequency of $f=1.24$~MHz in Figure \ref{fig:noisepsd}. In the overbiased state, the bolometer behaves as a resistor with noise contributions from Johnson noise as well as from the warm and cold readout electronics. When operated in the superconducting transition, the bolometer has an additional contribution from phonon noise \cite{dobbs2012}.   

The white noise noise level increases significantly between the overbiased and in-transition state, indicating that the contribution of the readout electronics is subdominant.  A few high-frequency lines are present in Figure \ref{fig:noisepsd}, occupying approximately 5\% of the total bandwidth.  These lines  are mostly outside the typical signal band ($<$30 Hz for a telescope with a one arcminute beam scanning at one degree per second).    Ideally, we would also evaluate the noise as a function of bias frequency.  However, the cryogenic components used with the room temperature electronics have known parasitic contributions that increase with frequency (see Section \ref{sec:cryocomponents}).  Therefore, in this work we focus only on the lower frequency bolometers ($<$2~MHz)  where the effect of these parasitics is less significant.  At these frequencies, we measure the readout noise to be essentially independent of frequency.  When the improved cryogenic components currently under development become available, we will evaluate the noise at higher frequencies.

\section{Conclusion}

We have presented room temperature electronics for the next generation digital fMux readout that multiplexes 64 bolometers together into a single SQUID module. These electronics include a SQUID controller, a mezzanine interface board, and a FPGA motherboard. The system bandwidth is increased from the legacy system to $\sim$10~MHz, allowing for additional $LCR_{\rm{bolo}}$ resonances. Digital Active Nulling (DAN) provides active feedback on the observed sky signal, reducing the dynamic range requirements of the SQUIDs. The digital synthesizer uses a novel polyphase filter bank approach to generate the carrier and nuller signals.

Flight representative versions of the SQUID controller and mezzanine boards were also developed. These boards were designed to operate in the environment of space, focusing in particular on high radiation tolerance and performance across a range of ambient temperatures. The boards were characterized in a temperature-controlled chamber from -$20^{\circ}$C to $+40^{\circ}$C.  They performed well at all temperatures, and a small benefit was found to operating below $0^{\circ}$C.  The boards were also designed to have low power consumption, a necessary criterion of space applications. For 64 detectors per readout module, we measure 49.8~mW per bolometer of power consumption for the flight representative electronics, which meets the LiteBIRD satellite design goal. Further power reductions are possible by increasing the multiplexing factor and by implementing the polyphase filter bank technique in the demodulator.

We demonstrate the ability of the new electronics to tune SQUIDs to their optimal operation point, bias detectors, and read out 64 channels per SQUID module. We have shown system performance using a test-module containing 44 bolometers and use the remaining 20 channels to characterize readout system noise.  The sub-Kelvin electronics used for these tests has stray reactances that become important at several MHz, so we focus on detectors at 1-2 MHz.  Parallel development and testing of new sub-Kelvin electronics is ongoing and these components will be available soon.
Comparing the white noise levels of the bolometers while overbiased and while operating in the superconducting transition, we find that the contribution of the readout electronics to the total measured white noise level is small compared to the intrinsic bolometer noise.

\section{Acknowledgments} 

The authors acknowledge funding from the Canadian Space Agency, the Natural Sciences and Engineering Research Council of Canada, the Canadian Institute for Advanced Research, and the Canada Research Chairs program.

We would like to thank the SPTpol collaboration for providing a detector wafer and LC board and NIST for providing the inductors and SQUIDS used in this work.  We thank Xilinx University Programs for their support.

\bibliography{spie_refs}   

\begin{thebibliography}{10}

\bibitem{benson2014}
{Benson}, B.~A. et~al., ``{The SPT-3G Experiment},'' in [{\em Society of
  Photo-Optical Instrumentation Engineers (SPIE) Conference
  Series}{\nolinebreak\hspace{0.1em}]},  {\em Society of Photo-Optical
  Instrumentation Engineers (SPIE) Conference Series} (2014).
\newblock In preparation.

\bibitem{tomaru2012}
{Tomaru}, T., {Hazumi}, M., {Lee}, A.~T., {Ade}, P., {Arnold}, K., {Barron},
  D., {Borrill}, J., {Chapman}, S., {Chinone}, Y., {Dobbs}, M., {Errard}, J.,
  {Fabbian}, G., {Ghribi}, A., {Grainger}, W., {Halverson}, N., {Hasegawa}, M.,
  {Hattori}, K., {Holzapfel}, W.~L., {Inoue}, Y., {Ishii}, S., {Kaneko}, Y.,
  {Keating}, B., {Kermish}, Z., {Kimura}, N., {Kisner}, T., {Kranz}, W.,
  {Matsuda}, F., {Matsumura}, T., {Morii}, H., {Myers}, M.~J., {Nishino}, H.,
  {Okamura}, T., {Quealy}, E., {Reichardt}, C.~L., {Richards}, P.~L., {Rosen},
  D., {Ross}, C., {Shimizu}, A., {Sholl}, M., {Siritanasak}, P., {Smith}, P.,
  {Stebor}, N., {Stompor}, R., {Suzuki}, A., {Suzuki}, J.-i., {Takada}, S.,
  {Tanaka}, K.-i., and {Zahn}, O., ``{The POLARBEAR-2 experiment},'' in [{\em
  Society of Photo-Optical Instrumentation Engineers (SPIE) Conference
  Series}{\nolinebreak\hspace{0.1em}]},  {\em Society of Photo-Optical
  Instrumentation Engineers (SPIE) Conference Series} {\bf 8452} (Sept. 2012).

\bibitem{austermann2012}
{Austermann}, J.~E., {Aird}, K.~A., {Beall}, J.~A., {Becker}, D., {Bender}, A.,
  {Benson}, B.~A., {Bleem}, L.~E., {Britton}, J., {Carlstrom}, J.~E., {Chang},
  C.~L., {Chiang}, H.~C., {Cho}, H.-M., {Crawford}, T.~M., {Crites}, A.~T.,
  {Datesman}, A., {de Haan}, T., {Dobbs}, M.~A., {George}, E.~M., {Halverson},
  N.~W., {Harrington}, N., {Henning}, J.~W., {Hilton}, G.~C., {Holder}, G.~P.,
  {Holzapfel}, W.~L., {Hoover}, S., {Huang}, N., {Hubmayr}, J., {Irwin}, K.~D.,
  {Keisler}, R., {Kennedy}, J., {Knox}, L., {Lee}, A.~T., {Leitch}, E., {Li},
  D., {Lueker}, M., {Marrone}, D.~P., {McMahon}, J.~J., {Mehl}, J., {Meyer},
  S.~S., {Montroy}, T.~E., {Natoli}, T., {Nibarger}, J.~P., {Niemack}, M.~D.,
  {Novosad}, V., {Padin}, S., {Pryke}, C., {Reichardt}, C.~L., {Ruhl}, J.~E.,
  {Saliwanchik}, B.~R., {Sayre}, J.~T., {Schaffer}, K.~K., {Shirokoff}, E.,
  {Stark}, A.~A., {Story}, K., {Vanderlinde}, K., {Vieira}, J.~D., {Wang}, G.,
  {Williamson}, R., {Yefremenko}, V., {Yoon}, K.~W., and {Zahn}, O., ``{SPTpol:
  an instrument for CMB polarization measurements with the South Pole
  Telescope},'' in [{\em Society of Photo-Optical Instrumentation Engineers
  (SPIE) Conference Series}{\nolinebreak\hspace{0.1em}]},  {\em Society of
  Photo-Optical Instrumentation Engineers (SPIE) Conference Series} {\bf 8452}
  (Sept. 2012).

\bibitem{kermish2012}
{Kermish}, Z.~D., {Ade}, P., {Anthony}, A., {Arnold}, K., {Barron}, D.,
  {Boettger}, D., {Borrill}, J., {Chapman}, S., {Chinone}, Y., {Dobbs}, M.~A.,
  {Errard}, J., {Fabbian}, G., {Flanigan}, D., {Fuller}, G., {Ghribi}, A.,
  {Grainger}, W., {Halverson}, N., {Hasegawa}, M., {Hattori}, K., {Hazumi}, M.,
  {Holzapfel}, W.~L., {Howard}, J., {Hyland}, P., {Jaffe}, A., {Keating}, B.,
  {Kisner}, T., {Lee}, A.~T., {Le Jeune}, M., {Linder}, E., {Lungu}, M.,
  {Matsuda}, F., {Matsumura}, T., {Meng}, X., {Miller}, N.~J., {Morii}, H.,
  {Moyerman}, S., {Myers}, M.~J., {Nishino}, H., {Paar}, H., {Quealy}, E.,
  {Reichardt}, C.~L., {Richards}, P.~L., {Ross}, C., {Shimizu}, A., {Shimon},
  M., {Shimmin}, C., {Sholl}, M., {Siritanasak}, P., {Spieler}, H., {Stebor},
  N., {Steinbach}, B., {Stompor}, R., {Suzuki}, A., {Tomaru}, T., {Tucker}, C.,
  and {Zahn}, O., ``{The POLARBEAR experiment},'' in [{\em Society of
  Photo-Optical Instrumentation Engineers (SPIE) Conference
  Series}{\nolinebreak\hspace{0.1em}]},  {\em Society of Photo-Optical
  Instrumentation Engineers (SPIE) Conference Series} {\bf 8452} (Sept. 2012).

\bibitem{grainger2008}
{Grainger}, W., {Aboobaker}, A.~M., {Ade}, P., {Aubin}, F., {Baccigalupi}, C.,
  {Bissonnette}, {\'E}., {Borrill}, J., {Dobbs}, M., {Hanany}, S.,
  {Hogen-Chin}, C., {Hubmayr}, J., {Jaffe}, A., {Johnson}, B., {Jones}, T.,
  {Klein}, J., {Korotkov}, A., {Leach}, S., {Lee}, A., {Levinson}, L., {Limon},
  M., {Macaluso}, J., {MacDermid}, K., {Matsumura}, T., {Meng}, X., {Miller},
  A., {Milligan}, M., {Pascale}, E., {Polsgrove}, D., {Ponthieu}, N.,
  {Reichborn-Kjennerud}, B., {Renbarger}, T., {Sagiv}, I., {Stivoli}, F.,
  {Stompor}, R., {Tran}, H., {Tucker}, G., {Vinokurov}, J., {Zaldarriaga}, M.,
  and {Zilic}, K., ``{EBEX: the E and B Experiment},'' in [{\em Society of
  Photo-Optical Instrumentation Engineers (SPIE) Conference
  Series}{\nolinebreak\hspace{0.1em}]},  {\em Society of Photo-Optical
  Instrumentation Engineers (SPIE) Conference Series} {\bf 7020} (Aug. 2008).

\bibitem{dobbs2012}
{Dobbs}, M.~A., {Lueker}, M., {Aird}, K.~A., {Bender}, A.~N., {Benson}, B.~A.,
  {Bleem}, L.~E., {Carlstrom}, J.~E., {Chang}, C.~L., {Cho}, H.-M., {Clarke},
  J., {Crawford}, T.~M., {Crites}, A.~T., {Flanigan}, D.~I., {de Haan}, T.,
  {George}, E.~M., {Halverson}, N.~W., {Holzapfel}, W.~L., {Hrubes}, J.~D.,
  {Johnson}, B.~R., {Joseph}, J., {Keisler}, R., {Kennedy}, J., {Kermish}, Z.,
  {Lanting}, T.~M., {Lee}, A.~T., {Leitch}, E.~M., {Luong-Van}, D., {McMahon},
  J.~J., {Mehl}, J., {Meyer}, S.~S., {Montroy}, T.~E., {Padin}, S., {Plagge},
  T., {Pryke}, C., {Richards}, P.~L., {Ruhl}, J.~E., {Schaffer}, K.~K.,
  {Schwan}, D., {Shirokoff}, E., {Spieler}, H.~G., {Staniszewski}, Z., {Stark},
  A.~A., {Vanderlinde}, K., {Vieira}, J.~D., {Vu}, C., {Westbrook}, B., and
  {Williamson}, R., ``{Frequency multiplexed superconducting quantum
  interference device readout of large bolometer arrays for cosmic microwave
  background measurements},'' {\em Review of Scientific Instruments}~{\bf 83},
  073113 (July 2012).

\bibitem{irwin1995}
Irwin, K.~D., ``An application of electrothermal feedback for high resolution
  cryogenic particle detection,'' {\em Applied Physics Letters}~{\bf 66}(15),
  1998--2000 (1995).

\bibitem{dehaan2012}
{de Haan}, T., {Smecher}, G., and {Dobbs}, M., ``{Improved performance of TES
  bolometers using digital feedback},'' in [{\em Society of Photo-Optical
  Instrumentation Engineers (SPIE) Conference
  Series}{\nolinebreak\hspace{0.1em}]},  {\em Society of Photo-Optical
  Instrumentation Engineers (SPIE) Conference Series} {\bf 8452} (Sept. 2012).

\bibitem{Lindeman2012}
Lindeman, M., Hijmering, R., Khosropanah, P., Korte, P., Hartog, R., Gottardi,
  L., Kuur, J., and Hoevers, H., ``Carrier phase optimization for frequency
  division multiplexing of low temperature detectors,'' {\em Journal of Low
  Temperature Physics}~{\bf 167}(5-6),  701--706 (2012).

\bibitem{smecher2012}
Smecher, G., Aubin, F., Bissonnette, E., Dobbs, M., Hyland, P., and MacDermid,
  K., ``A biasing and demodulation system for kilopixel tes bolometer arrays,''
  {\em Instrumentation and Measurement, IEEE Transactions on}~{\bf 61},
  251--260 (Jan 2012).

\bibitem{dobbs2008}
{Dobbs}, M., {Bissonnette}, E., and {Spieler}, H., ``{Digital Frequency Domain
  Multiplexer for Millimeter-Wavelength Telescopes},'' {\em IEEE Transactions
  on Nuclear Science}~{\bf 55},  21--26 (2008).

\bibitem{crochiere1983}
{Crochiere}, R.~E. and {Rabiner}, L.~R.,  [{\em Multirate Digital Signal
  Processing}{\nolinebreak\hspace{0.1em}]}, Prentice-Hall (1983).

\bibitem{smecher2012B}
Smecher, G., Aubin, F., George, E., de~Haan, T., Kennedy, J., and Dobbs, M.,
  ``An automatic control interface for network-accessible embedded
  instruments,'' {\em SIGBED Rev.}~{\bf 9},  23--27 (June 2012).

\bibitem{barron2014}
{Barron}, D. et~al., ``{Development and characterization of the readout system
  for POLARBEAR-2},'' in [{\em Society of Photo-Optical Instrumentation
  Engineers (SPIE) Conference Series}{\nolinebreak\hspace{0.1em}]},  {\em
  Society of Photo-Optical Instrumentation Engineers (SPIE) Conference Series}
  (2014).
\newblock In preparation.

\bibitem{hattori2014}
{Hattori}, K. et~al., ``{Frequency domain multiplexed readout of kilopixel
  arrays of transition edge sensor bolometers},'' in [{\em Society of
  Photo-Optical Instrumentation Engineers (SPIE) Conference
  Series}{\nolinebreak\hspace{0.1em}]},  {\em Society of Photo-Optical
  Instrumentation Engineers (SPIE) Conference Series} (2014).
\newblock In preparation.

\bibitem{henning2012}
{Henning}, J.~W., {Ade}, P., {Aird}, K.~A., {Austermann}, J.~E., {Beall},
  J.~A., {Becker}, D., {Benson}, B.~A., {Bleem}, L.~E., {Britton}, J.,
  {Carlstrom}, J.~E., {Chang}, C.~L., {Cho}, H.-M., {Crawford}, T.~M.,
  {Crites}, A.~T., {Datesman}, A., {de Haan}, T., {Dobbs}, M.~A., {Everett},
  W., {Ewall-Wice}, A., {George}, E.~M., {Halverson}, N.~W., {Harrington}, N.,
  {Hilton}, G.~C., {Holzapfel}, W.~L., {Hubmayr}, J., {Irwin}, K.~D.,
  {Karfunkle}, M., {Keisler}, R., {Kennedy}, J., {Lee}, A.~T., {Leitch}, E.,
  {Li}, D., {Lueker}, M., {Marrone}, D.~P., {McMahon}, J.~J., {Mehl}, J.,
  {Meyer}, S.~S., {Montgomery}, J., {Montroy}, T.~E., {Nagy}, J., {Natoli}, T.,
  {Nibarger}, J.~P., {Niemack}, M.~D., {Novosad}, V., {Padin}, S., {Pryke}, C.,
  {Reichardt}, C.~L., {Ruhl}, J.~E., {Saliwanchik}, B.~R., {Sayre}, J.~T.,
  {Schaffer}, K.~K., {Shirokoff}, E., {Story}, K., {Tucker}, C., {Vanderlinde},
  K., {Vieira}, J.~D., {Wang}, G., {Williamson}, R., {Yefremenko}, V., {Yoon},
  K.~W., and {Young}, E., ``{Feedhorn-coupled TES polarimeter camera modules at
  150 GHz for CMB polarization measurements with SPTpol},'' in [{\em Society of
  Photo-Optical Instrumentation Engineers (SPIE) Conference
  Series}{\nolinebreak\hspace{0.1em}]},  {\em Society of Photo-Optical
  Instrumentation Engineers (SPIE) Conference Series} {\bf 8452} (Sept. 2012).

\bibitem{huber2001}
Huber, M., Neil, P., Benson, R., Burns, D., Corey, A., Flynn, C.,
  Kitaygorodskaya, Y., Massihzadeh, O., Martinis, J.~M., and Hilton, G., ``Dc
  squid series array amplifiers with 120 mhz bandwidth (corrected),'' {\em
  Applied Superconductivity, IEEE Transactions on}~{\bf 11},  4048--4053 (Jun
  2001).

\bibitem{hazumi2012}
{Hazumi}, M., {Borrill}, J., {Chinone}, Y., {Dobbs}, M.~A., {Fuke}, H.,
  {Ghribi}, A., {Hasegawa}, M., {Hattori}, K., {Hattori}, M., {Holzapfel},
  W.~L., {Inoue}, Y., {Ishidoshiro}, K., {Ishino}, H., {Karatsu}, K.,
  {Katayama}, N., {Kawano}, I., {Kibayashi}, A., {Kibe}, Y., {Kimura}, N.,
  {Koga}, K., {Komatsu}, E., {Lee}, A.~T., {Matsuhara}, H., {Matsumura}, T.,
  {Mima}, S., {Mitsuda}, K., {Morii}, H., {Murayama}, S., {Nagai}, M.,
  {Nagata}, R., {Nakamura}, S., {Natsume}, K., {Nishino}, H., {Noda}, A.,
  {Noguchi}, T., {Ohta}, I., {Otani}, C., {Richards}, P.~L., {Sakai}, S.,
  {Sato}, N., {Sato}, Y., {Sekimoto}, Y., {Shimizu}, A., {Shinozaki}, K.,
  {Sugita}, H., {Suzuki}, A., {Suzuki}, T., {Tajima}, O., {Takada}, S.,
  {Takagi}, Y., {Takei}, Y., {Tomaru}, T., {Uzawa}, Y., {Watanabe}, H.,
  {Yamasaki}, N., {Yoshida}, M., {Yoshida}, T., and {Yotsumoto}, K.,
  ``{LiteBIRD: a small satellite for the study of B-mode polarization and
  inflation from cosmic background radiation detection},'' in [{\em Society of
  Photo-Optical Instrumentation Engineers (SPIE) Conference
  Series}{\nolinebreak\hspace{0.1em}]},  {\em Society of Photo-Optical
  Instrumentation Engineers (SPIE) Conference Series} {\bf 8442} (Sept. 2012).

\bibitem{khosropanah2012}
{Khosropanah}, P., {Hijmering}, R., {Ridder}, M., {Gao}, J.~R., {Morozov}, D.,
  {Mauskopf}, P.~D., {Trappe}, N., {O'Sullivan}, C., {Murphy}, A., {Griffin},
  D., {Goldie}, D., {Glowacka}, D., {Withington}, S., {Jackson}, B.~D.,
  {Audley}, M.~D., and {de Lange}, G., ``{TES arrays for the short wavelength
  band of the SAFARI instrument on SPICA},'' in [{\em Society of Photo-Optical
  Instrumentation Engineers (SPIE) Conference
  Series}{\nolinebreak\hspace{0.1em}]},  {\em Society of Photo-Optical
  Instrumentation Engineers (SPIE) Conference Series} {\bf 8452} (Sept. 2012).

\bibitem{litebird}
private communication (2011).

\bibitem{bock2009}
{Bock}, Philip, J.~J.~A., {Armus}, L., {Bally}, J., {Benford}, D., {Cooray},
  A., {Devlin}, M., {Dodelson}, S., {Dowell}, D., {Goldsmith}, P., {Golwala},
  S., {Hanany}, S., {Harwit}, M., {Holland}, W., {Holzapfel}, W., {Kenyon},
  {Matt}, {Irwin}, K., {Komatsu}, E., {Lange}, A., {Leisawitz}, D., {Lee}, A.,
  {Mason}, B., {Mather}, J., {Moseley}, H., {Meyer}, S., {Myers}, S., {Nguyen},
  H., {Novosad}, V., {Sadoulet}, B., {Stacey}, G., {Staggs}, S., {Richards},
  P., {Wilson}, G., {Yun}, M., and {Zmuidzinas}, J., ``{Superconducting
  Detector Arrays for Far-Infrared to mm-Wave Astrophysics},'' in [{\em
  astro2010: The Astronomy and Astrophysics Decadal
  Survey}{\nolinebreak\hspace{0.1em}]},  {\em Astronomy} {\bf 2010},  45
  (2009).

\end{thebibliography}
\bibliographystyle{spiebib}  
\end{document}